
 \input harvmac.tex
\Title{ \vbox{\baselineskip12pt\hbox{ YCTP-P2-93}\hbox{hep-th/9301121}}}
{New Modular Hopf Algebras related to rational $k$ $\widehat
{sl(2)}$}
\centerline{ Sanjaye Ramgoolam }
\centerline { Department of Physics}
\centerline { Yale University, New Haven CT 06511-8167}
\def\IN{\relax{\rm I\kern-.18em N}}
\noblackbox

\vskip .3in
   We show that the Hopf link invariants for an appropriate set of finite
dimensional representations of $ U_q SL(2)$ are identical, up to overall
normalisation, to the modular S matrix of Kac and Wakimoto for rational $k$
$\widehat {sl(2)}$ representations. We use this observation to construct new
modular Hopf algebras, for any root of unity $q=e^{-i\pi m/r}$, obtained by
taking appropriate quotients of $U_q SL(2)$, that give rise to 3-manifold
invariants according to the approach of Reshetikin and Turaev. The phase
factor correcting for the `framing anomaly' in these invariants is equal to
 $ e^{- {{i \pi} \over 4} ({ {3k} \over {k+2}})}$, an analytic continuation
of the anomaly at integer $k$. As expected, the Verlinde formula
gives fusion rule multiplicities in agreement with the modular Hopf algebras.
This leads to a proposal, for $(k+2)=r/m$ rational with an odd denominator,
for a set of $\widehat {sl(2)}$ representations obtained by dropping some
of the highest weight representations in the Kac-Wakimoto set and
replacing them with lowest weight representations. For this set of
representations the Verlinde formula gives non-negative integer fusion rule
multiplicities. We discuss the consistency of the truncation to highest and
lowest weight representations in conformal field theory.

\Date{1/93}
\newsec{Introduction}
  Integer $k$ $SU(2)$ WZW models of CFT have led to the discovery of some
intriguing relations between $\widehat {sl(2)}$ and $U_q SL(2)$. One notable
relation is the fact that the modular S matrix for the characters of the
integrable representations is proportional to the
Hopf link invariant computed from
knot theory, when we make a certain map between
representations of the two algebras.
Further, the 6j symbols of the quantum group are related to the braiding
matrices in WZW conformal field theory\ref\ms{ G. Moore
and N. Seiberg, Classical and Quantum Conformal field theory, Commun. Math.
Phys. 123, 177-254 (1989) }\ref\AGS{
L. Alvarez-Gaum\'e, C. Gomez, and G. Sierra, Duality and Quantum groups,
NPB330 (1990) 347 }.
 Since Kac and Wakimoto \ref\kawak { V. G. Kac and M. Wakimoto,
 Modular invariant
representations of infinite-dimensional Lie algebras and superalgebras,
Proc. Natl. Acad. Sci. USA, 88, 4956-4960, July 1988 } found a finite set of
representations of $\widehat {sl(2)}$  that close under modular
transformations, it has been expected that there
 might be a conformal field theory based on
these representations. Some properties of the characters and modular
 S matrix have
been studied in \ref\KoSo{I. G. Koh and P. Sorbah,
 Fusion Rules and sub-modular invariant
partition functions in
non-unitary theories, Phys. Lett. B215, 723-729}
\ref\mup{ S. Mukhi and S. Panda, Fractional-level current algebras
and the classification of characters, NPB 338(1990),
263-282}
\ref\walt{ P. Mathieu and M. A. Walton,
 Fractional level Kac-Moody algebras and Non Unitary Coset Conformal
Theories, Progress of Theor. Phys. Supplement, No. 102, 1990}.
How far do the connections between
 affine algebra and quantum
group go through at rational $k$ ?  What do they tell us about an eventual
CFT at rational $k$ and how much of the connections between 2D CFTs and 3D
gauge theories goes through at rational $k$ ?

 We find that it is possible to define a map between
 quantum group representations and
affine algebra representations such that the relation
 between modular S matrix and
Hopf link invariant is preserved.
 We find that new quotients of $U_q SL(2)$ can be defined, which are
 modular Hopf algebras.
These modular Hopf algebras allow the construction of 3-manifold invariants
which are characterised by a framing anomaly which indicates that they
might be related to a conformal field theory with $sl(2)$ current algebra
symmetry. The framing anomaly also
has a well defined large $k$ limit, which suggests that
they might be related to a 3D gauge theory.
 We show that the Verlinde formula gives fusion rule multiplicities in
agreement with the tensor products of the quantum group.

 For odd $m$, $k+2=r/m$, we can select out a set of
 highest weight representations of
 $\widehat {sl(2)}$ and their charge conjugates,  whose characters form a
representation of the modular group, so that the Verlinde
formula gives non negative integer fusion rule multiplicities.

Whereas the classical tensor products of highest weight with lowest weight
representations of $sl(2)$ can contain irreducible  continuous series type of
representations,
 we show that
 the quantum fusion of $\widehat {sl(2)}$ modules, one of which contains a
 certain null vector,
excludes the irreducible C-series. For $m=3$, any $r$, this null vector
condition suffices
to show that the entire set of highest and lowest weight
representations proposed
decouples from the irreducible continuous series.

\newsec{Definitions and notations}

In most of the considerations in this paper we will be concerned with
fusion rules, which follow from the null vector structure of the modules.
We will not be too concerned with unitarity and our discussion of
decoupling in section 10 will apply
whether one uses commutation relations in the $\widehat {su(2)}$ form or the
 $\widehat
 {su(1,1)}$ form.

  \subsec{ Definitions  for $\widehat {sl(2)}$}

The commutation relations of  $\widehat {su(2)}$ are the following
\eqn\comrels {\eqalign { &[J_m^0,J_n^{\pm}]=\pm
J_{m+n}^{\pm} \cr
& [J_m^0,J_n^0] = {{mk} \over 2} \delta_{m,-n}\cr
&[J_m^+,J_n^-]=2J_{m+n}^0 +
mk\delta_{m,-n}
\cr &[k,J_n^a]= 0,}}
where $a$ takes values $+,-,0$. For  unitary $\widehat {su(2)}$ modules
there is an inner product satisfying  the condition
$(J_n^+)^\dagger=J_{-n}^-$,
$(J_n^0)^\dagger =J_{-n}^0$.
 The commutation relations of $\widehat{
su(1,1)}$ are \ref\dpl{L. J. Dixon, M. E. Peskin and J. Lykken, Nucl. Phys.
B 325 (1989) 329 - 355 } \eqn\ncomrels{\eqalign{ &
 [J_m^0,J_n^{\pm}]=\pm J_{m+n}^{\pm} \cr
 & [J_m^0,J_n^0] = - {{mk}\over 2 } \delta_{m,-n}\cr &[J_m^+,J_n^-]
=-2J_{m+n}^0 + mk\delta_{m,-n},
\cr &[k,J_n^a]= 0,}}
Unitary modules have an inner product consistent with
 $(J_n^+)^\dagger=J_{-n}^-$,
$(J_n^0)^\dagger =J_{-n}^0$. Notice that taking $k$ to $-k$,
 and $J_{n}^+$ to
$-J_n^+$, converts the first set of commutation relations to the second.
Therefore for every highest weight module of compact  type, there is a map
to a highest weight module of the DPL type. For every compact module that
has a null vector this map gives us a DPL module that has a null vector.
And a formula for a null vector in the compact module can, by the simple
substitution give a formula for a null vector in the DPL module. For
example, $(J_{-1}^+)^R$ is a null vector in a highest weight module of
compact type, at any $k$,  when the highest weight $j$ is given by $2j+1=
-R+k+2$, and when $(J_{-1}^+)^R|j>$ is a null vector in a DPL highest weight
module with highest weight $j$ given by $2j+1= -R-k+2$. Another example is
the following \ref\sonnen{O. Aharony, O. Ganor, N. Sochen, J. Sonnenschein and
S.Yankielowicz,   Physical states in $G/G$ models and 2d Gravity,
 hepth-9204095 }. For any $k$, a highest weight module of
highest weight $j$ given by $j=-t/2, t=k+2$ has a null vector given by the
following formula
\eqn\chione {\chi =  [J_0^-J_{-1}^+J_0^- -t(J_0^-J_{-1}^0
+ J_0^-J_{-1}^0)-t^2J_{-1}^-]|j> .}
The same expression, with $t=-k+2$
and $J_{-1}^+$ replaced by $-J_{-1}^+$ gives a null vector in a DPL type
 module.
 This map does not necessarily give us a positive norm state if we start
with a
positive norm state, so if we require unitarity then these modules are very
different. We will work with $\widehat{ su(2)}$ notation in the rest of this
 paper and
since we will not be dealing with issues of unitarity, our conclusions
apply equally well if we had used $\widehat {su(1,1)}$ notation.

\subsec {  Definitions and conventions for $U_q sl(2)$ }
Recall the definition of $U_q
SL(2)$, which we will also call $U_q$
 \eqn \Uqsl { \eqalign{  [H,X_\pm] &= \pm X_\pm \cr
      [X_+ , X_-] &= {  { q^{2H}- q^{-2H} } \over { q-q^{-1}}  }  .\cr } }
The coproduct $ \Delta $, antipode $S$ and counit $\epsilon$ are :
\eqn\cac { \eqalign { \Delta (H) &= H \otimes 1 + 1 \otimes H \cr
                      \Delta (X_{\pm}) &= X_{\pm} \otimes q^H + q^{-H}
\otimes X_{\pm} \cr
 S(H) &= - H \cr
 S(X_+) &=- q X_+ \cr
 S(X_-) &= -q^{-1} X_-  \cr
 \epsilon (H) &=0 \cr
 \epsilon (X_{\pm}) &= 0 .\cr }}
One could also choose for the coproduct $ \Delta ' = \sigma \circ \Delta$,
where $$\sigma ( a\otimes b)= b \otimes a ,\qquad \forall a,b \in U_q $$
 At generic $q$, the universal R matrix \ref\kr{A. N. Kirillov,
N. Yu. Reshetikin, Representations of the algebra $U_qSl(2)$, $q$-orthogonal
polynomials and invariants of links. Leningrad, LOMI preprint E-9-88 1988 }
   satisfying $R \Delta = \Delta ' R$
 is given by
\eqn\rmat{\eqalign{  R &= q^{2 H \otimes H }
\sum_{n=0}^{\infty} q^{-1/2 n(n-1)}
{ {(1-q^{-2})}\over {[n]!} } q^{n(H\otimes 1 -1 \otimes H) }
(X_+)^n \otimes (X_-)^n  \cr
[n] &= { {q^n-q^{-n}} \over {q-q^{-1}} } \cr
[0]! &= 1 \cr
[n]! &= [n][n-1]...[1] \cr }}
At roots of unity, $q = e^{-i \pi m\over r} $ we can define
a quotient of $U_q sl(2)$, \ref\hub{ H. Saleur and J. B. Zuber,
Integrable lattice models and Quantum Groups, Lectures  given at the 1990
Trieste Spring School on String Theory and Quantum Gravity }\ref\keller{Georg
Keller, Fusion Rules of $U_q(sl(2,C))$, $q^m=1$, Lett. in Math. Phys.
21, 273-286,1991 }\ref\retu{N. Reshetikin
 and V. G. Turaev, Invariants of 3-manifolds via link polynomials and
quantum groups, Invent. math. 103, 547-597 (1991) }
by imposing the relations
$$(X_+)^r=0, (X_-)^r=0.$$ Then the sum in the R matrix truncates at $n=r$
\keller
\eqn\rmati{\eqalign{  R &= q^{2 H \otimes H }
\sum_{n=0}^{r-1} q^{-1/2 n(n-1)}
{ {(1-q^{-2})}\over {[n]!} } q^{n(H\otimes 1 -1 \otimes H) }
(X_+)^n \otimes (X_-)^n .  \cr}}
Notice that if we write K for $q^H$ and $K^{-1}$ for $ q^{-H} $
 we can still write down  the
coproducts, antipode and counit, for the smaller algebra generated by
$$ X_+, X_-, K, K^{-1} ,$$ instead of $H,X_+,X_-$. This algebra, called
$U_t$, is discussed by \retu .
If we call $e=q^H X_+$,
$f=q^{-H}X_-$ we can also write these down in terms of
 $$ e,f,K^2,K^{-2}$$ \ref\tjin{An introduction to quantized Lie Groups and
algebras, T. Tjin, Amsterdam preprint, (Nov 1991) }.
For any two representations $V_1$ and $V_2$ of these smaller algebras,
 the operator $R \in END(V_1 \otimes V_2)$ intertwines the two coproducts.
So with these generators we have all the structures
 of a quasitriangular Hopf algebra.
We will find it necessary to consider other algebras smaller than that
generated by $H,X_+,X_-$ but larger than that generated by
 $$e,f, K^2 ,K^{-2}. $$ Indeed we define for $$ q= e^{-i\pi m/r}, $$ the
algebra generated by $K^{2/m}, K^{-2/m}, e,f$ calling it
$U[m]$, equipped with
all the Hopf algebra structures inherited from $U_q$, for example
\eqn\crelm{ K^{2/m}eK^{-2/m} = q^{2H/m} q^H X_+ q^{2H/m} = q^2 q^HX_+ = qe .}

\subsec{ reducibility of Verma modules of $\widehat {sl(2)}$ and $U_q Sl(2)$ }

 We begin by drawing attention to a close relation between the existence of
null vectors in $sl(2)$ affine algebra Verma modules and the reducibility of
quantum group Verma modules for any $k \in \Re $. A highest weight
 representation of
$\widehat {sl(2)}$ with highest weight $j_i$ has null vectors when $j_i$ takes
 one of the
values parametrized by two integers $r_i$ and $s_i$
\eqn\affnull {\eqalign{ 2j_i +1 =
r_i - s_i (k+2)\quad &\hbox {with}\quad  r_i \in \IN ;\quad s_i \in \IN \cup
                                        \{0\} \cr
                     &\hbox{or} \quad -r_i \in \IN ; \quad -s_i \in \IN
.\cr} }
 The null vector
 occurs at
level $r_i s_i$ and has $J_0^0$ eigenvalue  $j_i - r_i $.
 This is a consequence of the Kac-Kazhdan
 determinant formula \ref\kacKaz{V. G. Kac and D. A. Kazhdan,
Structure of
Representations with Highest Weight of Infinite Dimensional Lie Algebras,
Advances in Mathematics, 34, 97 - 108 ( 1979) }. For generic $k$ a Verma
module has one null vector. For $k$ rational there are degenerate
representations with an infinite number of null vectors.

Using the formula
 \eqn\crel { [X_+,X_-^n] = [2H+n-1][n] X_-^{n-1 },  }
 which can be proved by induction, and writing
 $q = e ^ { - i\pi /\nu  },\quad \nu \in \Re $ we find that a $U_q$
 highest weight module with highest weight $j_i$ has a null vector
 with weight $ j_i - r_i $ for \eqn\qgred {  2j_i + 1
= r_i - s_i \nu , \hbox{ for } r_i \in Z_+ , \hbox{ and } s_i \in Z. }
The dimension of the module is $r_i$, and its quantum dimension $tr(q^{2H})$
is $e^{i\pi s_i} [r_i]$.

At generic $\nu$  there is one null vector in the Verma module with highest
weight $j_i$ of the form given above. Quotienting out the Verma module
by the submodule generated by the null vector, we obtain a finite
dimensional irreducible module of $U_q$. If we write $\nu$  as $(k+2)$
this formula is identical to \affnull,\ except for the ranges of $ r_i $ and
$s_i$.  For $s_i$ non-negative we can define
a map from the finite dimensional  irreducible quantum group
module with highest weight $j_i$ to the $\widehat {sl(2)}$
  irreducible module
with  highest weight $j_i$ (we are making here the natural identification
$ H \rightarrow J_0^0 $ ).
If the affine algebra module has a null vector
specified by $(r_i,s_i)$ with $ r_i, s_i \in Z_- $ then we will associate
with that highest weight representation   an irreducible quantum group
module of dimension $|r_i|$ and highest weight  $j_i+ |r_i|$.
 At rational $k$, we focus on highest weights that are in the Kac Wakimoto
set \kawak . They have the form
 \eqn \krat { 2j_i +1 = r_i - s_i (r/m),\qquad  r_i=1,2\cdots(r-1), \qquad
s_i= 0,1,2 \cdots (m-1).}  $j_i$ can also be written as $2j_i+1= (r_i-r)
- (s_i-m)(r/m) $. The module has two generating null vectors and we could
associate a quantum group module to it in two ways. Because of the
symmetries of $S$, either of these maps will have the property that the Hopf
link invariant is proportional to the modular transformation matrix. For
concreteness we will pick the map induced by the expression \krat.

In the next section we will calculate the regular isotopy invariant of the
Hopf link at rational $k$  and show that the map we have defined above
takes the Hopf link invariant to the modular  S matrix of Kac and
 Wakimoto up to an overall factor of $\sqrt {(2/rm)}$ . This
generalises the correspondence which is well known for
positive integer $k$ and which was exploited in \ref\witten {E. Witten
, Quantum Field Theory and the Jones Polynomial, Commun.
Math. Phys. 121, 351-399 (1989)}
 to construct 3 manifold
invariants from three dimensional Chern Simons field theory,
which are closely  related to modular functors defined in \ms.
Subsequently, Reshetikin  and Turaev gave a construction of 3D topological
field theories in the sense of Atiyah \ref\atiy{M. Atiyah, Topological
 Quantum Field
Theories, Publ. Math. IHES 68, 1988, 175 }
based on modular Hopf algebras.

\newsec  {Computation of the S matrix}

We will compute the regular isotopy invariant of the Hopf link
using the R matrix in the form
written down by Keller \keller\  for example. The same argument can be
carried through with  the expression of \retu .
\eqn\linkinv{  S_{Hopf}(j_1 , j_2) =
        (tr_{j_1} \otimes tr_{j_2} ) ( q^{2H} \otimes q^{2H} )
 (\check R)^2  }
Here $j_1$ and $j_2$ are highest weights of quantum group representations
$V_{1}$ and $V_{2}$.
$$ 2j_1 + 1 = r_1 - s_1 (r/m). $$
$$   2j_2 + 1 = r_2 - s_2 (r/m). $$
 $\check R $ is equal $ PR$ where P is the permutation operator.
It can be proved that as an operator acting on the space $V_1$ the
\eqn\onetrace{ (1 \otimes tr) (1 \otimes q^{2H}) (\check R)^2 }
is proportional to the identity. This is proved in \kr\
 using Schur's lemma,
the cyclicity
of the trace and the fact that
$$ \check R \Delta (a) = \Delta (a) \check R, \forall a \in U_q SL(2) $$
 So we act on the state $|j_1> \otimes
|m_2> $ , which is annihilated by all but the first term in \rmati , to get

\eqn\linkinvi {\eqalign { &( 1  \otimes tr)
 (1 \otimes q^{2H}) PRP R_{j_1 m_2}^{j_1 m_2}
 |j_1> \otimes |m_2>\cr
 &= (1 \otimes tr)(1 \otimes  q^{2H})
\sum_{n=0}^{r-1}   R_{m_2  j_1 }^{m_2 +n  j_1-n }
R_{j_1 m_2}^{j_1 m_2} |j_1 - n > \otimes |m_2 + n > \cr
&= \sum_{m_2 = j_2 - r_2 + 1 } ^ {j_2 } q^{2m_2} R_{m_2 j_1} ^ { m_2 j_1}
R_{j_1 m_2} ^ { j_1 m_2 } \cr
&= \sum_{ m_2 = j_2 - r_2 + 1} ^ {j_2} q^{2m_2} q^{4m_2j_1} \cr
&= e^{-{{2i \pi m}\over r} (j_2 + 1 - r_2)(2j_1 + 1 )}
 { {(1- e ^{-{{2 i \pi m}\over r} r_2  (2j_1 +1 )   } ) } \over
 {( 1 - e ^ {{{-2i\pi m}\over r} (2j_1+ 1)} ) }  }      \cr
&= e ^ {i \pi (s_1 r_2 + s_2 r_1 - s_1 s_2 r/m) }
 { {\sin (\pi m r_1 r_2/r)} \over  {\sin ( \pi m r_1  / r ) e^{i\pi s_1} }  }
.\cr }}
The Hopf link invariant is then obtained by multiplying this with
$$tr_1(q^{2H})= e^{i\pi
 s_1} \sin (\pi m r_1/r)$$ to get
\eqn\linkinvii { S_{Hopf}(j_1,j_2) = {1\over {\sin (\pi m/r)}}
 e^{i \pi (s_1 r_2 + s_2 r_1
 - s_1 s_2 r/m) }   \sin (\pi mr_1r_2/r). }
Now this is proportional to the Kac-Wakimoto modular S matrix, if we make
the identifications described in the previous section
\eqn\SandS { S_{Hopf}(j_1,j_2) =  { {\sqrt {(rm/2)} }
\over {\sin (\pi m/r)} }
S_{KW} (j_1,j_2) . }
For $m=1$ this equation is well known.

\newsec { Quantum group tensor products and the different algebras }
 We recall some facts about the tensor products of $U_q$ discussed in
detail in \keller\ . The quotient $U_t$ is discussed in \retu.
The complete set of highest weights $j_i$   of irreducible
representations   of $U_q$ with non zero $q$ dimension is given by
$$ 2j_i + 1 = r_i - s_i r/m  , \quad 1  \le r_i \le r-1, \quad s_i \in Z .$$
The tensor products of these
also contain  representations of zero $q$-dimension,  $I_{z} ^ {l}$ with
$z \in Z $
and  $ 1  \le l \le r  $. The representation with highest weight
characterised by the pair $ (r_i,s_i) $ will be denoted by $ <r_i,s_i> $.
Keller proves that the representations of zero $q$
dimension, $\{I_z^{l}| 1 \le l \le r,
z \in Z \}$ generate an ideal $I_z$  in the ring of representations $R_z$
of $U_q$.
This allows the definition of an {\it associative and commutative tensor
product }  $\hat {\otimes}$ on $R_z/I_z$.
With this tensor product we have
\eqn\tensprod { <r_1,s_1> \hat {\otimes} <r_2,s_2> = \bigoplus
_{r_3=|r_1-r_2|+1,+3,.. }^{min(r_1+r_2-1,2r-r_1-r_2-1)} <r_3, s_1 + s_2 > }
This set of representations has a {\it distinguished identity element} $<1,0>$
satisfying
\eqn\ide { <r_i,s_i> \hat {\otimes} <1,0> = <r_i,s_i>. }
For each representation $V_i$, also called $<r_i,s_i>$, there is a unique
$ V_{ \check i }$ or $<r_{\check i},s_{\check i}>$ with $r_{\check i} =
r_i$ and $ s_{\check i} = - s_i $ with the property that  the tensor
product of the two contains the identity representation.
$V_{\check i}$  is called the dual of $V_i$. Note that each representation
 has a
{\it unique dual, } and $$V_{ \check { (\check{i}) } } =
 V_i, \quad\hbox{for all
}i. $$
 One easily checks that the  dual of
 a representation of $U_q$ with highest weight $j_i$ has
 lowest weight $-j_i$.
 The set of irreducible highest weight representations of non zero $q$
dimension of the algebra $H,X_+,X_-$ is infinite because $s_i$ runs over
all integers. We will define quotients of this algebra for which the set of
non-isomorphic representations is finite.

 The following property of the representations $<r_i,s_i>$
allows us to deduce the properties of the other algebras we are
going to deal with, $U_t$ and $U[m]$ for $q=e^ {-i\pi m/r}$. It is easy
to prove \keller\
\eqn\kelma{ <r_i,s_i> \cong <r_i,0> \otimes <1,s_i> }
Now clearly  $<r_i,0>$ with distinct $r_i$ are non
isomorphic when regarded
as representations of any of the algebras $U_q$, $U_t$ or $U[m]$, because
they have dimension $r_i$. The
only state in $<1,s_i>$ is annihilated by $X_+$ and $X_-$, and $K^{2/m}$
has eigenvalue $ e^{( { {i\pi  s_i }\over {m} }) }$, while $K$ has
eigenvalue $ e^{( { {i \pi  s_i }\over 2 }) }$ . This means that the
residues of $s_i$  modulo $(2m)$  in case of $U[m]$ and the residues mod
$(4)$
in the case of $U_t$ suffice to determine the representations as
representations of
the respective algebras, up to isomorphism. We can thus deduce the tensor
products of representations for these different algebras from \tensprod .
The $s_3$ on
the right hand side gets replaced by $s_3\quad \hbox{mod}\quad (2m)$
 for $U[m]$.
The correspondence
with the description of the representations  of $U_t$ given in \retu\
is made by
noting that their fourth root of unity $\alpha $   can be identified with
$ e^{i\pi s_1/2} $.

We see that for $U[m]$, $U_t$ or $U_q$ there is a
commutative and associative tensor product, an identity representation
 and each representation
has a unique dual. It is also possible to write down the isomorphisms
between $V_{\check i}$ and $\check {V_i} $ with the
properties required by the
first two axioms of \retu, and to follow through the proof of the third
and fourth axioms, by using the above correspondence
between $U_t$, $U_q$ and $U[m]$.
\newsec{Modular Hopf Algebra }
  We will now show that, for each $(m,r)$, we
can pick a set of representations to make $U[m]$ a modular
 Hopf algebra as defined in
\retu . Reshetikin and Turaev give a general construction of three manifold
invariants and topological field theories in the sense of Atiyah \atiy\
from any modular Hopf algebra (MHA). They also give examples of MHA based on
$U_t$ for any odd  $m$, but in their modular Hopf algebras at odd $m$ the
framing anomaly does not have a well defined large $k$ limit, except for
$m=1$.
 We will use the algebra $U[m]$ and  our set of representations
 will be different from theirs, and will actually correspond (under the map
we have described in section 1) to a set of
representations closely related to the
Kac-Wakimoto representations, a relation which will
 be discussed in  more detail
in the section 9. The main interest of our new examples of
 modular Hopf algebras is
that
the  phase factor correcting for the framing anomaly
will be $ e^{- { {i\pi}\over 4} { {3k}\over{k+2} } } $
just as for positive integer $k$.
At large $k$, this becomes $e^{- { {i\pi}\over 4 }  (3) }$.
With a well defined large $k$ limit of the framing anomaly, it can be hoped
that these invariants could be understood as being related to 3 dimensional
Chern-Simons-Witten gauge theory. In the positive integer $k$ case the
large $k$ behaviour of the framing anomaly was part of the evidence
for the relation between the Reshetikin-Turaev construction and the Witten
invariants \retu, more detailed evidence is discussed in
\ref\frgo{ D. S. Freed and R. E. Gompf, Computer
 calculation of Witten's 3-manifold invariant,
 CMP 141, 79-117 (1991) } and references
therein.
 We will choose sets of quantum group representations
closed  under fusion and
satisfying the genus zero axioms \ms . And then we will check some genus
one equations which will prove that $U[m]$ is a modular Hopf algebra.
 We will prove
\smallskip {\bf Theorem 1}:
\eqn\stwo {    (S^2)_{i j}  = { {rm}\over 2 \sin^2(\pi m/r)}
 \delta_{i \check j} }
If we use the normalisation from affine algebra  this is equivalent to  :
\eqn\afstwo { (\tilde S ^2)_{i,j} = \delta_{i \check j} }
We will also prove
\smallskip {\bf Theorem 2}:
 \eqna\weights $$\eqalignno{
\hbox{If}\qquad \sum_{k} v_k S_{k,l} d_k
&=  v_l^{-1} dim_q (V_l) &\weights a \cr
 \hbox{then} \cr
 d_l &= \sqrt{2 \over {rm} } e^{-{{i\pi}\over 4}  {{3k}\over{(k+2)}} }
 \sin(\pi m/r) dim_q (V_l) &\weights b \cr
\hbox { which implies} \cr
 C &= \sum_{i } v_{i}^{-1}dim_q(V_{i})d_i =
e^{-{{i\pi}\over 2} {{3k}\over{(k+2)}} } \ne 0 &\weights c .\cr   }$$

The $v_i$  that appears above is the value of an element $v$ of $U_q$
in the representation of highest weight $j_i$ and is
 equal to $q^{-2j_i(j_i+1)}$
\AGS\retu.
In the construction of 3-manifold invariants in \retu\ the $d_l$ are weights
which enter the weighted averages of framed link invariants,  that are equal
to 3-manifold invariants. The equation \weights\ , together with the precise
form of the averages \retu,\   guarantees that these
averages are invariant under Kirby moves. These moves relate different links
for which surgery gives the same three manifold \ref\Kirby{R. Kirby, The
calculus of framed links in $S^3$, Invent. Math. 45,35-36(1978) }.

 Using $\tilde S $ the \weights a reads :
\eqn\afdl { (\tilde S)_{0l} = \sum_{k}  T_0 \tilde {S}_{0k} T_k \tilde
{S}_{kl}
 T_l,  }a special case of $ (\tilde S T)^3=C $,
 $C$ here being the charge conjugation
 matrix mapping a representation to its dual
 (distinct from the C in \weights ).
  The modular transformation
matrix on affine algebra characters $T_{jl} = T_j \delta_{j l} =
e^{ 2 \pi i[
 { {j(j+1)}\over{(k+2)} } - {c\over {24} } ] } $, $c=3k/(k+2)$.
 We note here the following properties of S,
 \eqna\props $$\eqalignno { \hbox{ in general }  S_{ij}  &\ne S_{i \check j}
            &\props a\cr
   \hbox{ but }  S_{\check i \check j} &= S_{i j} = S_{j i} &\props b\cr
\hbox {and in particular }
    S_{0 \check j}   &= S_{0 j} = S_{j0} &\props c .\cr} $$
  Note that from the
quantum group point of view the part of the T matrix
involving the classical casimir has a natural meaning but $c$ is only
determined after selecting an appropriate set of
representations and solving for the
weights.

We will discuss even $m$ and odd $m$ separately.

\newsec{ ODD $m$ }
We take as our set of representations those with
\eqn\repO{\eqalign{ r_i \in I(1) = \{1,2, \cdots , (r-1)\} \cr
                  s_i \in I(2) =\{0,\pm 2, \pm 4, \cdots , \pm (m-1)\}.\cr}}
These are all non isomorphic and have non zero q
dimension. The equation \tensprod\ and the moding by $2m$
discussed before implies that this set closes under the tensor product.
The indecomposables have zero $q$ dimension and we can define a
tensor product which is commutative and associative. The first four axioms of
\retu\ are satisfied. In the language of modular tensor categories
described in \ref \rcftn {G. Moore and N. Seiberg, Lectures on Rational
 Conformal Field Theories, Lectures given at Trieste Spring School 1989 and
Banff Summer School 1989 }
\ms, these guarantee that the genus zero axioms are
satisfied.

\subsec{ computation of $S^2$ }
We have for the square of the S matrix
 \eqn\ssqu {\eqalign{  (S^2)_{j_1 j_3}  &= { 1 \over {\sin^2(\pi m/r)} }
\sum_{r_2,s_2} e^{i\pi[r_2(s_1+s_3)+ r_2(s_1 + s_3) ]}
e^{-i\pi [s_2 r (s_1+s_3)/m] }\cr
 &\qquad \qquad \sin (\pi m r_1 r_2/r) \sin ( \pi m r_2 r_3/r ) \cr
 &={ 1 \over {\sin^2(\pi m/r)} }
 \sum_{s_2} e^{-i\pi s_2 	r(s_1 + s_3)/m}
    \sum_{r_2}    \sin (\pi m r_1 r_2/r) \sin ( \pi m r_2 r_3/r ). \cr }}
The $s_2$ sum  can be written as
$$ e^{2i\pi(m-1)r/m } { {[1- e^{-2i  \pi r (s_1 + s_3 )}]} \over
{[1- e^{-2\pi i r  (s_1 + s_3)/m }]} }. $$
 Note that the numerator is always zero whereas the denominator is only non
zero if $(s_1 +s_3)$  is zero or  a multiple of $m$. Since  $s_i$ are all
even, no two of them can sum up to $\pm m$  which is odd, and any higher
multiple is not possible because of the range. So the $s_2$ sum is only non
zero if $s_1=-s_3$ and then the sum is $m$. Now  the summand in the $r_2$
sum is symmetric under change of sign of $r_2$ and it is zero when $r_2= 0$
or $r$. So we can write it as
 $$ 1/2 \sum_{r_2=(-r+1),(-r+2)..}^r
 \sin (\pi m r_1 r_2/r) \sin ( \pi m r_2 r_3/r )$$
After writing in this form and doing the four geometric series we
see that the sums will vanish unless $\pm (r_1 \pm r_3)=0, \pm 2r,\cdots$.
 Now
the ranges of $r_1$ and $r_3$ guarantee that the only possibility is
 $r_1=r_3$, when the sum is $ r/2 $. This proves equation \stwo .

\subsec{ computation of $d_i$ }
 Using \stwo\ we see that \weights a \ implies that
\eqn\stwowts {\eqalign {& q^{-2 j_{\check k}( j_{\check k} + 1 ) }
d(\check r_k, \check s_k) rm/2   \cr
&= \sum_{r_i \in I(1)} \sum_{s_i \in I(2)} q^{2j_i(j_i+1)}
\sin(\pi m r_i/r) \sin(\pi m r_i r_k/r)e^{i\pi[r_is_k+ r_ks_i -
s_is_kr/m)]}\cr
&= \sum_{r_i \in I(1)} \sum_{s_i \in I(2)} e^{-i\pi m/ (2r)
[r_i^2 - 1  + s_i^2r^2/m^2 ] } e^{-i\pi[s_is_k r/m]} \sin (\pi mr_i/r)
  \sin(\pi m r_i r_k/r)\cr
&= \sum_{s_i \in I(2)} e^{-i \pi r/(2m)(s_i^2 + 2s_is_k)}
\cr
& \sum_{r_i \in I(1)} \sin (\pi mr_i/r) \sin (\pi mr_ir_k/r)
e^{-i\pi m/2r(r_i^2 - 1)}\cr }}
If we write $ s_i' = s_i/2$ in the $s_i$ sum. Note that $s_i'$
runs over all residues modulo $m$, and the exponential only depends on
residue class modulo $m$, so we can shift $ s_i'$ without changing
the sum. So the $s_i$ sum is equal to
\eqn\ssumi  { e^{i\pi (r/2m) s_k^2} \sum_{ s_i'(m) }
e^{-i\pi(2r/m) {s_i'}^2 } ,}
where we have adopted the notation $s_i'(m)$ for a sum over $s_i'$ running
over a complete set of residues modulo  ($m$).
The result for the $r_i$ sum can be extracted from \retu\
\eqn\orsumi{\eqalign{ &\sum_{r_i=1}^{r-1} \sin (\pi mr_i/r)
\sin (\pi mr_ir_k/r)
e^{-i \pi m/2r(r_i^2 - 1)}\cr
&= {{e^{3i\pi m/2r -i\pi/2}}\over 2}
e^{{{i\pi m} \over {2r}} (r_k^2-1)}  \sin(\pi mr_k/r)
\sum_{k=0}^{2r-1} e^{-i \pi k^2m/(2r)} }}
A short manipulation then shows that
\eqn\manip { d(r_k,-s_k) = {1\over{(rm)}} e^{-i\pi/2 + 3i\pi
m/(2r)} \sin(\pi mr_k/r)\sum_{ s_i' (m)} e^{-i\pi(2r/m){
s_i' } ^2 } \sum_{k(2r)} e^{-i\pi k^2m/(2r)}   }
This product of Gaussian sums equals another Gaussian sum
$$ \sum_{h (2rm)} e^{-i\pi h^2/(2mr)}= \sqrt{2rm}e^{-i\pi/4}. $$
 This can be proved by imitating
the steps in the proof of a similar multiplicative formula for Gauss sums
given  in \ref\hardwr{ G. H. Hardy and G. M. Wright, An introduction to the
theory of numbers, Clarendon Press, 1989}. The result then is
\eqn\weitod{ d(r_k,s_k) =d( \check r_k,\check s_k)=
 \sqrt{ {2\over{rm}} }e^{-{{3i\pi}\over 4} { {(r-2m)} \over r}}
\sin(\pi mr_k/r) }
 This proves theorem 2.

\newsec{EVEN $m$}
The set of irreducible representations of $U[m]$ we choose, have
highest weights $j_i$
characterised by
\eqn\repsE{\eqalign{ r_i \in I(1) &= \{1,3,5...r-2\} \cr
\hbox{ and}\quad  s_i \in I(2) &= \{0, \pm 1, \pm 2,....\pm (m-1), m \}
\cr}}
 The set of representations closes under the tensor product.
 We prove that with this choice of representations we
have theorems 1 and 2 .
\eqn\ssquare {\eqalign {
  (S^2)_{j_1 j_3}  &=  { 1 \over {\sin^2(\pi m/r)} }
\sum_{r_2,s_2} e^{i\pi[r_2(s_1+s_3)
+ s_2(r_1 + r_3) ]}e^{-i\pi [s_2r/m (s_1+s_3)]}\cr
&\qquad \qquad \sin (\pi m r_1 r_2/r) \sin ( \pi m r_2 r_3/r ) \cr
 &={ 1 \over {\sin^2(\pi m/r)} }
 \sum_{s_2\in I(2)} e^{-i\pi s_2 r/m(s_1 + s_3)}
    \sum_{r_2\in I(1)} e^{i\pi r_2(s_1+s_3)}   \cr
&\qquad \qquad \sin (\pi m r_1 r_2/r) \sin ( \pi m r_2 r_3/r ) }}
We used $e^{i\pi s_2 (r_1+r_3)}=1$ because $r_1$ and $r_3$ are both odd.
 The $s_2$ sum  can be written as
$$ e^{2i\pi(m-1)r/m } { {[1- e^{-2i  \pi r (s_1 + s_3 )}]} \over
{[1- e^{-i  \pi r (s_1 + s_3 )/m}]} } .$$
Now the numerator is zero but the denominator is also zero if $(s_1 + s_3)$
is $0\quad \hbox{mod}\quad (2m)$. Now from the range of $s_i$ it is
 clear that $0$ is the
only possibility for any $s_1 \ne m$, and for $s_1=m$, $s_3=m$ is the only
possibility.
So the $s_2$ sum is equal
 to $2m \delta (s_1, -s_3 \quad \hbox{mod} \quad 2m)$. Then
the $s_1,s_3$ dependent phase drops out of $r_2$ sum.
\eqn\sumr {\eqalign{    &\sum_{r_2 = 1,3..r-2}
                   \sin (\pi m r_1 r_2/r) \sin ( \pi m r_2 r_3/r ) \cr
& =1/2  \sum_{r_2'= 0, \pm 1, \pm 3 .. \pm (r-2) }
                   \sin (\pi m r_1 r_2/r) \sin ( \pi m r_2 r_3/r ) \cr }}
Again we expand the sines to get  geometric sums of the form
$$ \sum_{r_2= 0, \pm 1, \pm 3 .. \pm (r-2) } e^{i\pi m Nr_2/r} $$
 proportional to
$$   { {1-e^ {2i\pi mN}} \over
{1-e^ {2i\pi r_2mN/r}  } }   $$
where N can be $\pm (r_1 \pm r_3) $. The sum is non vanishing only if $N=
0$ (mod $r$). Now because of the range of $r_i$ negative multiples
 and positive
multiples higher than the first are clearly excluded. Also
$r$ is odd because
we required $(r,m)=1$. $r_1 + r_3$ cannot be zero because they are both
positive and it cannot be $r$ because they are both odd and must add to an
even number. For the same reason $r_1-r_3$ cannot be  $r$ so the only non
zero contribution arise from $r_1= r_3$, so that the $r_2$ sum is
${r\over 4}
\delta (r_1,r_3)$. This proves \stwo\ for even $m$.

 Now we consider the determination of the weights.
 Using \stwo\ we see that \weights a implies that
\eqn\stwowtse {\eqalign{ &q^{-2 j_{\check k}( j_{\check k} + 1 )}
d(\check r_k,\check s_k) rm/2   \cr
 &= \sum_{r_i \in I(1)} \sum_{s_i \in I(2)} q^{ 2 j_i(j_i+1) }
\sin(\pi m r_i/r) \sin(\pi m r_i r_k/r)\cr
 &\qquad \qquad e^{i\pi[r_is_k+ r_ks_i -
s_is_kr/m)]} e^{i\pi s_i} \cr
&=e^{i \pi s_k}  \sum_{s_i \in I(2)} e^{- i \pi r/(2m)
(s_i^2 + 2s_i(s_k+m) )}
\cr
&\sum_{r_i \in I(1)} \sin (\pi mr_i/r) \sin (\pi mr_ir_k/r)
e^{-i\pi m/2r(r_i^2 - 1)} \cr }}
where we have used the fact that $e^{i\pi r_is_k}= e^{i\pi s_k}$ for
$r_i$ odd.
Now  $e^{-i \pi r s_i^2/(2m)}$ only depends on residue class of $s_i$
 mod $(2m)$ . This allows a shift in the $s_i$ sum to give
$$   e^{ { {i\pi(s_k+m)^2}\over{2m} } } \sum_{ s_i' (2m) }
e^{-i\pi(r/2m) {s_i'}^2 } $$
 The $r_i$  sum can be written as
\eqn\ersumi{\eqalign{ &\sum_{r_i=1,3.. (r-2) } \sin (\pi mr_i/r)
 \sin (\pi mr_ir_k/r)
e^{-i \pi m(r_i^2 - 1)/2r} \cr
&= \sum_{r_i=1,3.. (r) } \sin (\pi mr_i/r) \sin (\pi mr_ir_k/r)
e^{- i \pi m(r_i^2 - 1)/(2r)} \cr
&= {{e^{-i\pi/2}}\over 2} e^{i\pi m[r_k^2+2]/(2r)} \sin(\pi mr_k/r)
\sum_{r_i=1,3..2r-1} e^{-i \pi (m/2r) r_i'^2} .\cr  }}
The first equality above is a simple observation that the summand at
$r_i=r$ is zero, the second uses some steps very similar to those used in
\retu\  in their computation of the weights.
 We then obtain the following equation
\eqn\wegt{\eqalign{  d(r_k,-s_k)&= {{e^{-i\pi[1/2 - s_k]}}\over {rm} }
e^{+3i\pi (m/2r)} e^{i \pi rm/2}
 \sin(\pi mr_k/r) \cr
&\sum_{r_i =
1,3..}^{2r-1}  e^ {-i\pi (m/2r)r_i'^2}
 \sum_{s_i (2m)} e^{-i\pi (r/2m) s_i^2} }}
$r_i'$  runs over all the odd residues mod (2r) so we add the even number
$r+1$ to it without changing the sum to get  $$
\sum_{r_i =
1,3..}^{2r-1} e^{-i\pi (2m/r)r_i'^2}
= e^{-i\pi mr/2} \sum_{r_i''=0,1..r-1} e^{-i\pi (2m/r)r_i''^2}, $$
where we also changed the summation index to $r_i''=(r_i'+1)/2$.
$$ {{e^{i \pi[-1/2+s_k]}}\over{rm}} e^{3i\pi m/2r} \sin(\pi mr_k/r)
 \sum_{r_i''(r)} e^{-i\pi (2m/r)r_i''^2} \sum_{s_i (2m)}
e^{-i\pi(r/2m)s_i^2} $$
The two Gauss sums can again be combined into a Gauss sum over residues mod
$(2mr)$ \hardwr , which can be evaluated. The result is then
\eqn\wts{ d(r_k,s_k) = d(\check r_k,\check s_k)=
\sqrt {{2\over {rm}}} e^{{{-3i\pi}\over 4} (r-2m)/r}
\sin(\pi mr_k/r) e^{i\pi s_k},   } proving theorem 2 for the even $m$ case.

\newsec{Fusion rules and Verlinde formula}
We show in this section that the fusion rules as computed using the
Verlinde formula \ref \ver { E. Verlinde,
Fusion Rules and Modular transformations in 2D
CFT, Nucl. Phys. B300 (1988) 360 }
\ref\poleq{G. Moore and N. Seiberg,
Polynomial equations for Rational Conformal field Theories,
 Phys. Lett. 212B (1988) 451 } agree with
 the tensor product structure of representations of $U[m]$.
 In particular there are no minus signs in $N_{ij}^k$ .
 We will use $\tilde S$, with the normalisation natural from the
relation to $\widehat {sl(2)}$. We will prove :
\eqn\verform { \eqalign{N_{ij}^k &= \sum_{n} {\tilde S}_{in} {\tilde S}_{j n}
 {\tilde S}_{k n}^*/{\tilde S}_{0n}\cr
&= \Delta (r_i,r_j;r_k) \delta (s_k, s_i+s_j \hbox{mod} \quad 2m)\cr }}
where
\eqn\Delt {\eqalign{ \Delta (r_i,r_j;r_k) &= 1 \cr
                             \hbox{if}\quad r_k&\in \{|r_i - r_j|+1,...
min(r_i+r_j-1,2r-r_i-r_j-1)\} \cr
                                          &=0 \quad\hbox{otherwise} \cr}}
This is exactly the fusion of the Hopf algebra $U[m]$. Note that
\eqn\tens{  N_i = N_i (sl(2)_{r-2}) \otimes N_i (U(1)_{m/2}) ,} where we
are using the notation of \rcftn\ for the abelian fusion rules.

\subsec{ ODD $m$}
Writing out the formula we have
\eqna\oddN $$\eqalignno{  N_{ij}^k &=
  \sum_{r_n s_n } {\tilde S}(r_i s_i;r_n s_n)
{\tilde S}( r_j s_j; r_n s_n)
 {\tilde S}^*(r_k s_k; r_n s_n ) / {\tilde S}(1,0 ;r_n s_n) &\oddN a\cr
        &= ({2\over{rm}})\sum_{r_n s_n} e^{i \pi s_n(r_i+ r_j -r_k-1)}
e^{ i \pi r_n( s_i +
s_j-s_k) } e^{ i\pi s_n ( s_i + s_j - s_k )r/m } \cr
        &\qquad \qquad \sin (\pi m r_ir_n/r )\sin ( \pi m r_j r_n/r)
{{\sin ( \pi m r_k r_n/r)}\over {\sin ( \pi m r_n/r)}} &\oddN b\cr
       &=({2 \over {rm}})\sum_{s_n=0,\pm 2,.. \pm (m-1)}
 e^{ i\pi s_n ( s_i + s_j - s_k )r/m} \cr
& \sum_{r_n=1,2..(r-1)}
\sin (\pi m r_ir_n/r )\sin ( \pi m r_j r_n/r)\cr
& \qquad \qquad \qquad \sin ( \pi m r_k r_n/r)
/\sin ( \pi m r_n/r). &\oddN c\cr} $$
The first two exponentials in \oddN b are equal to one because the $s_i$
are chosen to be even.
Now the $s_n$ sum is $m \delta ( s_i+s_j-s_k, 0\quad\hbox{mod}\quad 2m)$,
 since $s_i+s_j-s_k$ cannot equal to $m$ which is odd. Using the symmetry
under change of sign of $r_n$, and noting that the summand is zero for
$r_n=0 ,r$, we  rewrite the $r_n$ sum as
$$ 1/2 \sum_{r_n=0, \pm 1, \pm 2.. \pm (r-1), r}
 \sin (\pi m r_ir_n/r )\sin ( \pi m r_j r_n/r)\sin ( \pi m r_k r_n/r)
/\sin ( \pi m r_n/r).$$
 Now $r_i$ is running over a complete set of residues modulo $2r$ and the
sum depends only on residue class mod $(2r)$. But if $r_n$ runs over a
complete set of residues mod $(2r)$, so does $mr_n$ because $(m,2r)=1$. So
this $r_n$  sum is the same as at $m=1$. This sum can be written down
because we know the fusion rules for $k$ positive integer. It is
$ {r\over 2}  \Delta (r_i,r_j:r_k) $. The claim is then
proved for odd $m$.

\subsec{EVEN $m$}
\eqn\evenN {\eqalign{  N_{ij}^k &=  \sum_{r_n s_n } {\tilde S}(r_i s_i;r_n s_n)
{\tilde S}( r_j s_j; r_n s_n)
 {\tilde S}^*(r_k s_k; r_n s_n ) / {\tilde S}(1,0 ;r_n s_n) \cr
         &= {2\over {rm}}\sum_{r_n s_n } e^{i \pi s_n(r_i+ r_j -r_k-1)}
  e^{ i \pi r_n( s_i + s_j-s_k)} e^{ i\pi s_n ( s_i + s_j - s_k )r/m}\cr
         &\sin (\pi m r_ir_n/r )\sin ( \pi m r_j r_n/r)\sin ( \pi m r_k r_n/r)
/\sin ( \pi m r_n/r) \cr  }}
Now  $$e^{i \pi s_n(r_i+ r_j -r_k-1)}=1$$ because the $r_i$ are all odd.
The $s_n$ sum is then seen to equal $  2m \delta (s_i + s_j - s_k, 0 \quad
\hbox{mod}\quad 2m)$.
This condition for the product of sums to be non zero guarantees that the
exponential appearing in the $r_n$ sum is also equal to 1. Then we are left
with
$$  \sum_{r_n=1,3..(r-2)}
 \sin (\pi m r_ir_n/r )\sin ( \pi m r_j r_n/r)\sin ( \pi m r_k r_n/r)
/\sin ( \pi m r_n/r) $$
Using the symmetry
under change of sign of $r_n$, and noting that the summand is zero for
$r_n= r$, we  rewrite the $r_n$ sum as
$$ 1/2 \sum_{r_n= \pm 1, \pm 3.. \pm (r-2), r}
 \sin (\pi m r_i r_n/r )\sin ( \pi m r_j r_n/r)\sin ( \pi m r_k r_n/r)
   /\sin ( \pi m r_n/r) $$
 Now $r_i$ is running over a complete set of odd residues modulo $r$ and the
sum depends only on residue class mod $(r)$ . Now if $r_n$ runs over a
complete set of residues mod $(r)$, so does $m r_n$ because $(m,r)=1$.
By the
same argument as for odd $m$ case then the sum is
$ {r\over 4}  \Delta (r_i,r_j,r_k) $. The claim is then
proved for even $m$.

\newsec{ Relation to $\hat sl(2)$}
  So we have chosen a set of quantum group representations which could
consistently describe the braiding and modular properties of a conformal
field theory. We chose a set that closely resembled the Kac Wakimoto set,
under the map we described between $U_q$
and $\widehat {sl(2)}$. We can use the
map now to propose, for odd $m$, a set of $\widehat {sl(2)}$
representations closely
related to
the Kac-Wakimoto set.
This set contains both highest weight and lowest weight
representations, since we
want a set where every representation has a dual.
We will then prove, directly using relations between the characters of
highest weight and lowest weight representations properties
  that the modular invariant
 partition functions written
in \KoSo\
 can also be interpreted as partition
functions for this alternative spectrum. Then we will discuss some other
possibilities for the state space of some CFTs with $sl(2,R)$
current algebra symmetry.

For odd $m$, we take highest weight representations
 with highest weights $j_i$ given by
\eqn\hiwe{ j_i(h.w)=  { (r_i-1)\over 2} -{s_i \over 2}(r/m) }
 with $r_i=1,2 \dots r-1$ and $ s_i=0, 2,  4 \dots
(m-1) $. And we take lowest weights
\eqn\lw{ j_i(l.w) = -j_i(h.w), \quad \hbox{ for }  s_i \ne
0. }
 The $s_i=0$ representations are self dual and are only counted once.

This set has exactly the same number of representations as the Kac-Wakimoto
set but some of the
highest weights there have been replaced by lowest weights. We have to make
sure that this set also has all the characters necessary to give a
representation of
the modular group.  This is guaranteed by the relations between the
characters of the highest weight modules that we
 have dropped and the characters
of the lowest weight modules that we have included.

It can be
shown using properties of the characters \KoSo\ that
\eqn\chare{ \chi_ {D^-(r_i,s_i)} (\tau,z) = -\chi_{D^-(r-r_i,m-s_i)}(\tau,-z)
,\qquad  s_i \ne 0.}
We can also prove the following
\smallskip {\bf Proposition }:
\eqn\redplmi{  \chi_{D^-(j_i)}(\tau,z) =
 \chi_{D^+(-j_i)}(\tau ,-z).}
where $D^+(j)$ is an irreducible lowest weight representation
 of lowest weight $j$ and
$D^-(j)$ is an irreducible highest weight
representation of highest weight $j$.
\smallskip{\bf Proof.}
   We first prove the analogous result for Verma modules $V^-_{j}$ and
$V^+_{-j}$ then use the fact that
the characters of the irreducible modules are obtained as an alternating
 sum of characters of Verma modules.
\eqn\dplmi { \eqalign{ \chi_{V^-(j)}(\tau,z) &=
 Tr_{V^-(j)} [ e^{2\pi i (\tau L_0 -
 z J_0^0)} ] \cr
   &= {{e^{ 2 \pi i [{ {\tau j(j+1)} \over {(k+2)}} -zj]} } \over  {1-e^{2\pi
iz} }   } \cr
    \qquad  &\prod_{n = 1}^\infty  {1\over {(1-e^{2\pi i \tau n})(1-e^{2 \pi i
\tau n} e^{2 \pi i z}) (1-e^{2 \pi i \tau n} e^{-2 \pi i z})}  }  \cr}}
We have used the fact that the monomials
\eqn\basmin{ \prod_{i=1}^\infty (J_{-i}^+)^{p_i,+} \prod_{i=1}^{\infty}
(J_{-i}^0)^{p_{i,0}}\prod_{i=0}^\infty (J_{-i}^-)^{p_{i,-}}  |j> }
form a basis for the Verma module. This is a consequence of the
 Poincare-Birkhoff-Witt theorem as explained
 for example in \ref\BS{M. Bauer and N.
Sochen, Fusion and
Singular vectors in $A^{(1)}_1$ highest weight cyclic modules, Hepth.
/9201079, Saclay Prep., SphT/91-117 }.
For the lowest weight module  a basis is given by
\eqn\baspl{ \prod_{i=1}^\infty (J_{-i}^-)^{p_i,-} \prod_{i=1}^{\infty}
(J_{-i}^0)^{p_{i,0}}\prod_{i=0}^\infty (J_{-i}^+)^{p_{i,+}}  |-j>   }
Clearly then the character is
\eqn\dplmii {\eqalign{ \chi_{V^+(-j)}(\tau,z) &= Tr_{V^+(-j)}
 [ e^{2\pi i (\tau L_0 - zJ_0^0)} ] \cr
   &= { {e^{ 2 \pi i [{ {\tau j(j+1)} \over {(k+2)} } + zj] } }
 \over  {1-e^{-2\pi
iz} }   } \cr
    \qquad  &\prod_{n = 1}^\infty  {1\over {(1-e^{2\pi i \tau n})(1-e^{2 \pi i
\tau n} e^{2 \pi i z}) (1-e^{2 \pi i \tau n} e^{-2 \pi i z})}  }  \cr}}
The change in sign of the exponent in the denominator comes about because
of the change in the range of summation in  \baspl\ compared to \basmin .
This shows that
\eqn\verm{\chi_{V^-(j)}(\tau,z) = \chi_{V^+(-j)}(\tau,-z) }
The same formula clearly relates characters
of highest and lowest weight modules
related by change of sign of $J_0^0$ eigenvalue of the distinguished
vector, for any level $n$. The characters look like the ones above except that
the $L_0$ eigenvalue is shifted by $n$.
But the characters of the  irreducible modules are equal to alternating
sums of characters of Verma modules \kawak \ref\BF{
D. Bernard and G. Felder, Fock space representations
and BRST cohomology in $sl(2)$ current algebra,
Commun. Math. Phys. 127 , 145-168 (1990)}. The Verma modules in the expression
for the character of $D^+(-j)$ are related to those in the expression for
the character of $D^-(j)$ by change of sign of the $J_0^0$ eigenvalue of
the distinguished vector. This is guaranteed by an automorphism of the
algebra $\widehat {sl(2)}$. Together with \verm\  this completes the
proof of \redplmi.

In the special case of integrable modules this equation reduces to the
statement that an integrable module with highest weight $j$, which can be
regarded as an irreducible quotient of a highest weight module with
highest weight $j$ or as an irreducible quotient of a lowest weight
module with lowest weight $-j$, is self-dual.

 Equations \chare\ and \redplmi\ imply that
\eqn\repl{\eqalign{ -\chi_{D^-(j_i')}(\tau,z) &=
\chi_{D^+(-j_i)}(\tau , z)\cr
 \hbox {where} \quad 2j_i+1 &= r_i - s_i(r/m) \cr
\hbox {and} \quad   2j_i'+1  &= (r-r_i) - (m-s_i)r/m \cr }}
 This means that the set of characters in \hiwe\ and \lw\
differs from the Kac-Wakimoto set by having some characters replaced by
their negatives. Clearly then we can write modular invariant partition
functions for this spectrum. That the S matrix for this set of
representations gives
non-negative
integer fusion rules follows from the computations in section 8, together
with the relation
\eqn\rels{ S(r_1,s_1;r-r_2,m-s_2) = - S(r_1,s_1;r_2,-s_2). }

The choices of representations that are consistent with
 modular invariant partition
functions are not unique. In particular if $\chi_i$ is replaced by
$\chi_i + C_{ij}\chi_j$, the equations  $SC=CS$ and $SS^{\dagger}=1$
 guarantee
that we get other modular invariant partition functions where the state
space of the theory is modified by replacing the field $\phi_i$ in the
chiral part with the
direct sum of $\phi_i$ with its charge conjugate.

 For even $m$, although we have a modular Hopf algebra our set of
representations
includes highest weights with $s_i=m$ which do not appear in the Kac
Wakimoto set. While the modular S matrix and the characters
certainly possess analytic  continuations to these values, and the Hopf
link invariant can still be computed, we do not know the characters and
modular transformation properties for $s_i$ outside the Kac wakimoto set.

\newsec{Decoupling of the irreducible continuous series}

 Let us consider the set of highest weight representations that has
 a null vector
 of the form $(J_{-1}^+)^R$ and deduce the most general consequences for the
fusion of these representations arising from setting the null vector to zero.
The currents are :
\eqn\cur{J^a(z) = \sum _{n=-\infty}^\infty J_n^a z^{-n-1} }

We define
primary fields by their operator products \ref\kz{
V. G. Knizhnik and A. B. Zamolodchikov, Current algebra and
 Wess-Zumino model in two dimensions, NPB247(1984)
83-103 }
with the currents
\eqn\prima { J^+(z) \phi ^j_m (w) =
 { {\tau ^+ \phi ^j_{m}(w) } \over{(z-w)} } + \hbox{Regular
terms}. }
\eqn\primb { J^0(z) \phi ^j_m (w) =
{ {\tau ^0 \phi ^j_{m}(w)}\over{(z-w)} } + \hbox{Regular
terms}. }
\eqn\primc { J^-(z) \phi ^j_m (w) = { {\tau ^- \phi ^j_{m}(w)}\over
 {(z-w)} }
+ \hbox{Regular terms}.  }
  The $\tau$ 's are operators representing the zero mode $sl(2,R)$ subalgebra.
The label m, not necessarily integral, is equal to the eigenvalue
of the compact generator $J_0^0$. It
runs in integer steps  over $(-\infty,\infty)$ in the case of the continuous
series, over a semi-infinite interval for the discrete series type.
  Choosing to consider irreducible modules means that we should set null
vectors to zero. Therefore
 \eqn\nul{<[(J_{-1}^+)^R\phi^{j_1}_{j_1}(z_1)]
\phi^{j_2}_{m_2}(z_2)\phi{j_3}_{m_
3}(z_3)>=0.}
where the action of $J_{-n}^a$ \ref\gepwit{ D. Gepner and E. Witten,
String theory on group
manifolds, NPB278 (1986) 493-549} is defined by,
\eqn\opdef{J_{-n}^a\phi^{j_1}_{m_1}(z_1)=\int dw{{J^a(w)}\over{(w-z_1)^n}}
 \phi^{j_1}_{m_1}(z_1),}
the $w$ integral being taken round a small contour around $z_1$.
\eqn\movecont{\eqalign{&<\int dw {{J^+(w)}\over {w-z_1}} \phi^{j_{1}}_{j_1}
(z_1)\phi^{j_{2}}
_{m_2}(z_2)\phi^{j_{3}}_{m_3}(z_3)> \cr
&= < \int {{dw}\over{w-z_1}}
 \phi^{j_{1}}_{j_1}(z_1)
 J^+(w)  \phi^{j_{2}}_{m_2}(z_2)\phi^{j_{3}}_{m_3}(z_3)>
\cr &\quad +< \int {{dw}\over{w-z_1}}
 \phi^{j_{1}}_{j_1}(z_1)
   \phi^{j_{2}}_{m_2}(z_2) J^+(w)  \phi^{j_{3}}_{m_3}(z_3)>
 \cr &= < \int {{dw}\over{w-z_1}}
 \phi^{j_{1}}_{j_1}(z_1)
 {{\tau^+_2}\over{w-z_2}}  \phi^{j_{2}}_{m_2}(z_2)\phi^{j_{3}}_{m_3}(z_3)>
\cr  &\quad +< \int {{dw}\over{w-z_1}}
 \phi^{j_{1}}_{j_1}(z_1)
   \phi^{j_{2}}_{m_2}(z_2) {{\tau^+_3}\over{w-z_3}}
\phi^{j_{3}}_{m_3}(z_3)> \cr
& =<
 \phi^{j_{1}}_{j_1}(z_1)
 {{\tau^+_2}\over{z_1-z_2}}
\phi^{j_{2}}_{m_2}(z_2)\phi^{j_{3}}_{m_3}(z_3)> \cr
 &\quad +<
 \phi^{j_{1}}_{j_1}(z_1)
   \phi^{j_{2}}_{m_2}(z_2) {{\tau^+_3}\over{z_1-z_3}}
\phi^{j_{3}}_{m_3}(z_3)>
.\cr}} Iterating this argument R times we obtain a sum of partitions of
the $\tau^+$ operators  acting at $z_2$ and $z_3$. We can isolate the term
where all the $\tau ^ +$ are acting at $z_2$ by its singularity ${1\over
{(z_1 -z_2)^R}}$, to get
\eqn\selrule{ 0= <\phi^{j_1}_{j_1}(z_1)(\tau^+)^R
\phi^{j_2}_{m_2}(z_2)\phi^{j_3}_{m_3}>.}
 This means that any $\phi^{j_2}_{m_2'}$ which can be written as
$(\tau^+)^R\phi^{j_2}_{m_2}$ will vanish in a three point function with
$\phi^{j_1}_{j_1}$ if the highest weight module generated by
$\phi^{j_1}$ contains the null vector, i.e if $2j_1+1= -R+(k+2)$.
Any state in the modules of
 irreducible continuous series type is of this form.
Now all the states in this module are generated by the action of the
generators on this highest weight, so the correlation function of any state
in this module with the $\phi^{j_2}_{m_2'}$ is zero. The proof goes as
follows. Let Y be an element of the enveloping algebra of $\widehat {sl(2)}$
,generated by $J_0^-$ and $J_{-1}^+$. Then
$<[Y\phi^{j_1}_{j_1}(z_1)]\phi^{j_2}_{m_2}(z_2)\phi^{j_3}_{m_3}(z_3)>  $ can
be equated by the above contour deformation argument to a sum of terms of
the form $<\phi^{j_1}_{j_1}(z_1)[{\hat Y_2}]\phi^{j_2}_{m_2}(z_2)]
[{\hat Y_3}\phi^{j_3}_{m_3}(z_3)]>$ multiplied by some powers of $(z_1-z_2)$
and $(z_1-z_3)$, where ${\hat Y}$ is some element generated by $J_0^+$ and
$J_0^-$.
By the above argument these all vanish. Again using the action of the
 generators and contour
deformation shows that all the states in the
 irreducible continuous series module
labelled by $j_2$ do decouple  in correlation functions
from the states in the $j_1$ module.

  The same null vector  for $j_1=0$, $k$ positive integer, was used in
\gepwit\
using the  insertion of the identity operator to show that all the
  correlation functions of fields transforming according to nonintegrable
representations vanish. If we are looking at other values of $k$, then the
$R$
cannot be
a positive integer and still satisfy $2j_1+1= -R+k+2$ for $j_1$ equal to
zero. So the correlation functions for infinite dimensional $sl(2)$
representations at
 the base are not necessarily zero.

  The simple null vector used above also further constrains the fusion of
highest weight representations. If $\phi^{j_2}$ belongs to a highest weight
representation then
any three point function including it and $\phi^{j_1}$ will vanish.
Correlation functions of $\phi^{j_1}$ with
lowest weight representations are not
forced to be zero.
This is probably related to the `trivialisation' of the OPE discussed by
Dotsenko in \ref\dotse {V.I.S. Dotsenko, Solving
 the $su(2)$ CFT using the Wakimoto free field
representation, NPB358, 1991, 547 }.
 We expect that the argument for the decoupling of the irreducible
continuous series
representations can be generalised to any $k$, for any highest
or lowest weight module
containing any null vector.

The condition $2j_1+1 = -R + (k+2) = -R + r/m $ can be rewritten as
\eqn\ratk{ 2j_1 + 1 = (r-R) - (m-1)r/m .}
For $R<r$ this belongs to the set of highest weights in the Kac Wakimoto
set. For $m = 3$ $s_i= 2= m-1$ is the only non zero value of $s_i$. These,
we have proved, decouple from the irreducible continuous series.
 This means that for
$k+2= r/3$ for any $r$, the simple null vector argument guarantees that the
entire set we are proposing decouples from the irreducible continuous series.

\newsec{CONCLUSIONS }
The maps we have used between affine algebra modules and quantum groups may
appear a little stranger than the known maps familiar at positive integer
$k$
where affine algebra modules with a finite base are mapped to quantum group
modules with the same dimension as the base. We note that Bauer and Sochen
\BS\ in their
construction of null vectors in highest weight modules
 define an action of the subalgebra generated by
$J_{-1}^+, J_0^-$ on a finite set
of vectors starting from the highest weight and leading to the null vector.
For $r_i$ positive we pick the action of $J^+_{s_i}$, and by use of
 the commutation
relations define an action of $J_{-s_i}^-$ and $(J_0^0- s_i k/2)$. This
gives an $r_i$
dimensional representation of $Sl(2)$ starting
 at the highest weight $j_i$. The
$U_q Sl(2)$ we associate with each generating null vector of a highest
weight Verma module is perhaps usefully thought of as a deformation of such
an action of $Sl(2)$.

We have found no natural explanation from the quantum group point of view
of the choice of $U[m]$. We arrived at it by exploiting the
correspondence of S matrices, but it would be interesting if this choice of
a quotient of $U_q$ could be understood directly from the quantum group.

 It will be interesting to understand the relation of the
 3-Manifold invariants to 3D gauge theories. In fact, after completing this
 paper,
we received a paper of C. Imbimbo \ref\imb{C. Imbimbo, New modular
representations and fusion algebras from quantized $SL(2,R)$
 Chern-Simons Theories, GEF-TH 2/1993, hepth-9301031} where
the fusion rules in equation \tens\ are arrived at from $SL(2,R)$ Chern-Simons
theory.
 Another problem is to find
 how to realise the fusion rules and braiding matrices in  conformal
field theory.

\bigbreak\bigskip\bigskip\centerline{{\bf Acknowledgements}}\nobreak
I would like to thank Gregory Moore for many discussions, help and guidance.
I would like to thank Hubert Saleur for  helpful discussions about quantum
groups and knot invariants, and Ronen Plesser for many useful discussions. I
would also like to thank James Horne, Wai Ming Koo and Jintai Ding for useful
discussions.

\listrefs
\end